\documentclass{article}
\usepackage{spconf,amsmath,graphicx}
\usepackage[sort,numbers]{natbib}

\usepackage{amsmath,amsfonts,bm}

\def\1{\bm{1}}
\newcommand{\train}{\mathcal{D}}

\def\sp{space}

\def\rvx{{\mathbf{x}}}
\def\rvy{{\mathbf{y}}}

\DeclareMathAlphabet{\mathsfit}{\encodingdefault}{\sfdefault}{m}{sl}
\SetMathAlphabet{\mathsfit}{bold}{\encodingdefault}{\sfdefault}{bx}{n}

\def\gY{{\mathcal{Y}}}

\newcommand{\vsp}{\csname v\sp \endcsname}
\newcommand{\hsp}{\csname h\sp \endcsname}

\usepackage{symbols}
\usepackage[usenames]{color}                                                        
\definecolor{mycitecolor}{rgb}{0,0.08,0.45} 
\usepackage[colorlinks=true, citecolor=mycitecolor]{hyperref}
\usepackage[export]{adjustbox}
\usepackage{makecell}
\usepackage{multirow}

\newcommand*{\ShowNotes}{} 
\definecolor{darkred}{rgb}{0.7,0.1,0.1}
\definecolor{darkgreen}{rgb}{0.1,0.7,0.1}
\definecolor{cyan}{rgb}{0.7,0.0,0.7}
\definecolor{dblue}{rgb}{0.2,0.2,0.8}
\definecolor{maroon}{rgb}{0.76,.13,.28}
\definecolor{burntorange}{rgb}{0.81,.33,0}
\definecolor{tealblue}{rgb}{0.212,0.459, 0.533}
\ifdefined\ShowNotes
  \newcommand{\colornote}[3]{{\color{#1}\bf{#2: #3}\normalfont}}
\else
  \newcommand{\colornote}[3]{}
\fi

\title{A Comparison Study on Infant-Parent Voice Diarization}
\name{Junzhe Zhu, Mark Hasegawa-Johnson and Nancy McElwain\thanks{Thanks to Jiahao Xu from University of Sydney for help with server. This work was supported by funding from the National Institute on Drug Abuse (R34DA050256-01), the National Institute of Mental Health (R21MH112578-01) and the National Institute of Food and Agriculture, U.S. Department of Agriculture (ILLU-793-339).}}
\address{University of Illinois at Urbana-Champaign}

\begin{document}
%
\maketitle
\begin{abstract}
We design a framework for studying prelinguistic child voice
from 3 to 24 months based on state-of-the-art algorithms in diarization. Our system consists of a time-invariant feature extractor, a context-dependent embedding generator, and a classifier. We study the effect of swapping out different components of the system, as well as changing loss function, to find the best performance. We also present a multiple-instance learning technique that allows us to pre-train our parameters on larger datasets with coarser segment boundary labels. We found that our best system achieved 43.8\% DER on test dataset, compared to 55.4\% DER achieved by LENA software. We also found that using convolutional feature extractor instead of logmel features significantly increases the performance of neural diarization.
\end{abstract}

\section{Introduction}
\label{sec: intro}

Mental health disorders and behavioral problems first emerge in early
childhood~\cite{egger2006common},~\cite{Cree2018}.
Early diagnosis and intervention may help to ameliorate or prevent some types of behavioral disorders, but findings for the effectiveness of interventions are often mixed~\cite{Wright2016}. Intensive assessments of parent-infant interactions in naturalistic home environments, as well as normative data, are needed. Yet, such assessments conducted manually pose logistical  challenges, including time and labor required by researchers, as well as privacy concerns of participating families. Automatic or semi-automatic diarization has the potential to address these limitations and provide insight into parent-infant vocal interactions -- including the relative timing, duration, volume, and tone of voice -- and would permit the establishment of normative data while minimizing privacy concerns. Greater volume of normative data, in turn, would facilitate the creation of effective evidence- based interventions in support of child mental health~\cite{Hoagwood2001}.



\begin{figure*}[t]
\centering
\includegraphics[width=0.9\linewidth]{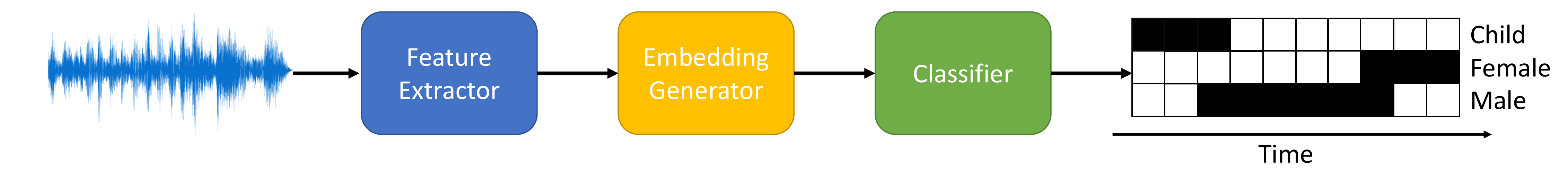}
\caption{An overview of system components. Feature extractor, embedding generator, and classifier are successively applied to input waveform, yielding a prediction for each speaker class at each frame.}
\label{fig: system}
\end{figure*}

\label{sec: related}
In contrast to broadcast news recordings~\cite{Canseco-Rodriguez2004,Tranter06}, automatic diarization of naturalistic parent-child recordings is difficult because: (1) most speakers are usually recorded at a distance, (2) most utterances are informal and brief, (3) infants’ voices and types of utterances differ significantly from those of adults. Diarization of distant
recordings~\cite{Pardo2006,Stolcke2010} and informal
speech~\cite{Castaldo2008} has been well-studied, and end-to-end
neural solutions~\cite{Fujita2019EndtoEndNS} exist.  Automatic diarization of children's speech, however, has only been extensively studied in the past three or four years, and remains challenging.
Perhaps the most influential child speech diarization system is LENA~\cite{Canault2015LENA,Cristia2018TalkerDI}, a propriety system that segments audio recorded from LENA wearable devices, and that has been commonly used as a baseline~\cite{Franc2018TheAD,Cristia2018TalkerDI}. 
Additional recent work on
child speech diarization has been inspired by 
the Second DiHARD Challenge~\cite{Ryant2019TheSD}, which includes childrens' speech (the Seedlings Corpus~\cite{Bergelson12916}) as one of its test corpora. 

Because of the difficulty of the task, most systems submitting results to DiHARD use oracle voice activity detection  (VAD)~\cite{8462311,Zajc2018ZCUNTISSD,Xie2019MultiPLDADO}; other solutions in the literature
include automatic VAD~\cite{Franc2018TheAD,Garca2019SpeakerDI} or
explicit models of one or more non-speech classes~\cite{7846253,Cristia2018TalkerDI,lavechin2020opensource}.
When oracle VAD is not provided, it is not always clear how a system should respond.
Parent-infant interaction involves overlap of speech events, interspersed by long periods of silence. 
Instead of assigning each frame to a single type, it is more natural to formulate the problem as a multi-label sequence-to-sequence problem \cite{lavechin2020opensource,Garca2019SpeakerDI,Fujita2019EndtoEndNS}.
Furthermore, the permutation-invariant~\cite{Yu2017PermutationIT} labelling rules typical of other diarization tasks are less appropriate for infant databases, in which there  is typically one infant, one female adult, one male adult, and sometimes one other child of a different age. Infant diarization accuracy may improve by pre-training and/or clustering models of 3 or 4 classes, e.g., {key child, other child, female adult, male adult}~\cite{Franc2018TheAD,Cristia2018TalkerDI,Franc2018TheAD,7846253,Garca2019SpeakerDI,lavechin2020opensource}.

This paper demonstrates end-to-end neural diarization of infant vocalizations. We also present a way to pre-train the neural network on large datasets with inaccurate boundary labels. In addition, we study the effect of different loss functions and input segment lengths of the network.  Sec.~\ref{sec: system} is the system description, Sec.~\ref{sec: pretrain} describes our approach to pre-training, Sec.~\ref{sec: experiment} describes experimental methods and results, Sec.~\ref{sec: conclusions} concludes.

\section{System Description}
\label{sec: system}

Suppose we have a dataset $\cD = \{(\rvx, \gY)\}$ where $\rvx \in \mathbb{R}^T$ is a waveform of $T$ samples, and $\gY \in \mathbb{R}^{C \times L}$ is the label. $C$ is the number of speaker classes that we define, and $L$ is the number of frames, where $\rvy_{c, l}$ is the binary label indicating whether any speaker from class $c$ is speaking at any sample in the range of $\Big[\frac{l}{L}T, \frac{l+1}{L}T\Big)$. We wish to design a system $F_{\theta}$ parameterized by $\theta$ that approximates the true mapping 
\be
\rvx \xrightarrow[]{F_{\text{true}}} \gY
\label{eq:x2y}
\ee
from the audio to binary speaker class labels for each frame.

We decompose our system as follows:
\be
F_{\theta}(\rvx) = (\text{Sigmoid} \circ F_{\text{cls}} \circ F_{\text{embed}} \circ F_{\text{feat}})(\rvx)
\ee
where 
\be
F_{\text{feat}} : \mathbb{R}^T \xrightarrow[]{} \mathbb{R}^{H \times L}
\ee
is a time-invariant mapping from signal samples to feature space, 
\be
F_{\text{embed}} : \mathbb{R}^{H \times L} \xrightarrow[]{} \mathbb{R}^{E \times L}
\ee
maps each frame from feature space to speaker embedding, and is conditioned on the rest of the frames. Finally, 
\be
F_{\text{cls}} : \mathbb{R}^{E \times L} \xrightarrow[]{} \mathbb{R}^{C \times L} \equiv \mathbb{R}^{E} \xrightarrow[]{} \mathbb{R}^{C}
\ee
maps each frame in the embedding sequence to a set of logits, and does not depend on other frames.

\label{ssection: feature}
{\bf{Features $F_\text{feat}$:}} Two types of features are tested.

The first tested feature vector is based on a 23-dimensional log-Mel-filterbanks with a window size of 25ms and hop size of 16ms. We splice features of 15 consecutive frames, and subsequently subsample the spliced feature matrix by a factor of 16 along the time dimension. Therefore, we compute a 345-dimensional feature vector every 256ms.

The second tested feature vector consists of a learned filterbank applied to waveform samples.
This method is based on the encoder in \cite{Stoller2018WaveUNetAM}, and consists of 12 blocks of Conv1D with zero-padding, followed by LeakyReLU activation, then Decimation Pooling which halves the time dimension. At a waveform sampling rate of 16000Hz, this feature extractor produces a 288-dimensional feature vector every 256 ms.

\label{ssection: embedding}
{\bf{Embedding $F_\text{embed}$}:} Two types of neural embeddings were tested.

The first tested neural embedding is a Bi-Directional LSTM (BLSTM)~\cite{10.1162/neco.1997.9.8.1735}. Similar to \cite{Fujita2019EndtoEndNS}, we use 5 layers with 256 hidden units for both forward and backward LSTM. 

The second tested neural embedding is a multi-headed self-attention model.  As in~\cite{Fujita2019EndtoEndNS}, we linearly transform each frame's feature vector, then apply two encoder layers.  As in~\cite{Vaswani2017AttentionIA}, each encoder layer includes  layer normalization~\cite{Ba2016LayerN}, then multi-headed self-attention, then another layer norm, then a two-layer position-wise fully-connected network. After both encoder layers, another layer normalization is applied. For self attention, we set both input and output dimensions to 256, and for the position-wise fully-connected network, we set the hidden layer size to 1024.

\label{ssection: classifier}
{\bf{Classifier $F_\text{cls}$:}}
To map the speaker embedding of each frame to a binary label for each speaker class, we use either a linear predictor, or a two layer fully connected network with ReLU activation (denoted as MLP), with the first hidden layer the same size as speaker embedding.

\label{ssection: loss}
{\bf{Focal Loss for Imbalanced Binary Labels:}}
We use Adam~\cite{Kingma2015AdamAM} to train our networks.  The majority of frames are silence: without considering voice overlap, for a label tensor of size $\mathbb{R} ^ {C \times T}$, only $\frac{T_{\text{on}}}{C \times T}$ of target entries will be $1$, where $T_{\text{on}}$ is the total number of frames where someone is speaking. Therefore, we treat this as a class-imbalanced classification problem, and use focal loss~\cite{Lin2017FocalLF} to balance our training. As with the best configuration in \cite{Lin2017FocalLF}, we use $\alpha=0.25, \gamma=2$ as our hyper-parameters. When running on test data, we set the prediction threshold to 0.5.

\begin{figure}[t]
\begin{minipage}[b]{.48\linewidth}
  \centering
  \centerline{\includegraphics[width=3.0cm]{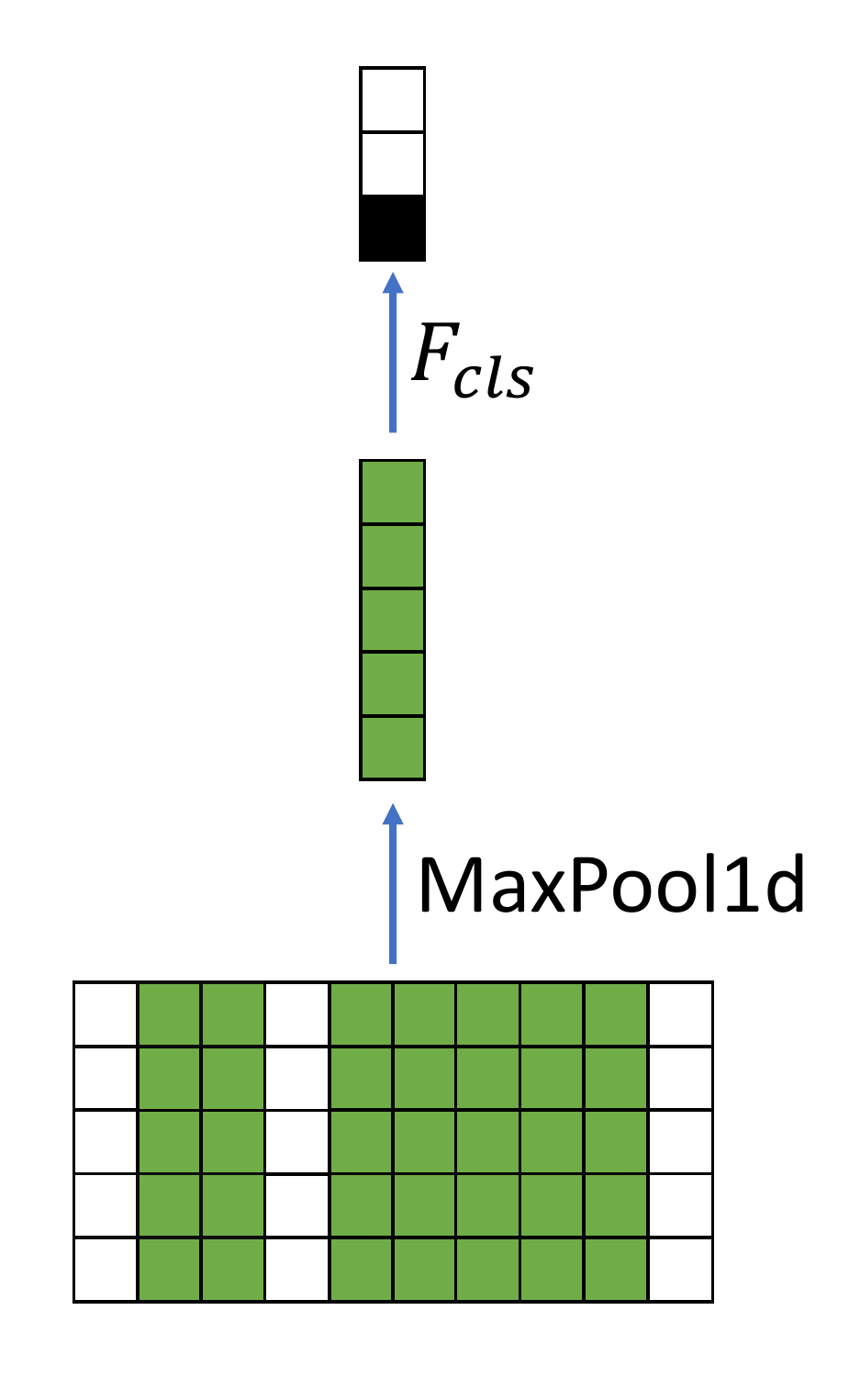}}
  \centerline{(a) MIL1}\medskip
\end{minipage}
\hfill
\begin{minipage}[b]{0.48\linewidth}
  \centering
  \centerline{\includegraphics[width=3.0cm]{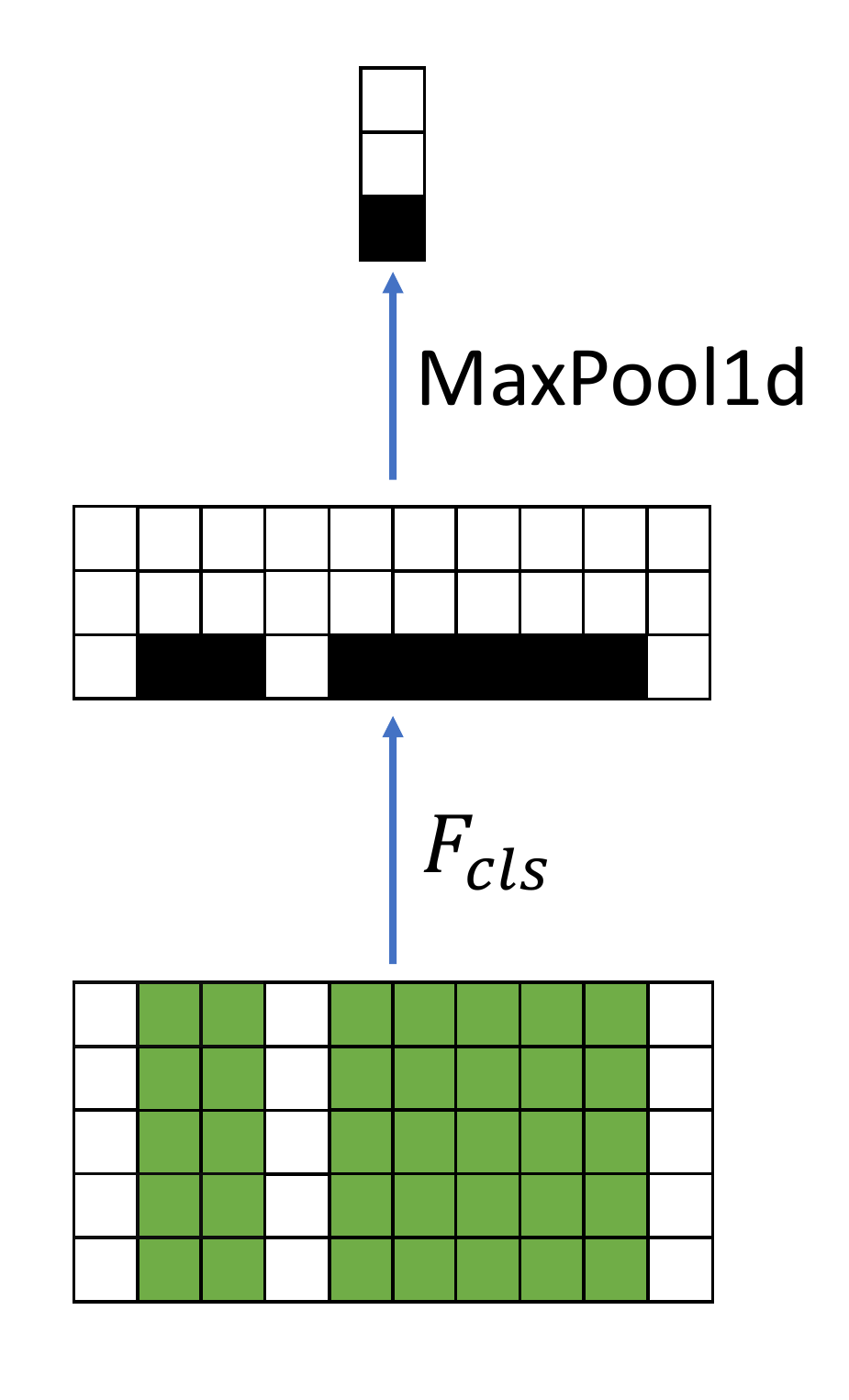}}
  \centerline{(b) MIL2}\medskip
\end{minipage}
\caption{Two configurations for multiple-instance learning. Colored blocks are speaker embedding. Black and white boxes are speaker class.}
\label{fig:res}
\end{figure}

\begin{table*}[t]
\centering
\begin{tabular}{| c | c | c | c | c | c | c | c | c | c | c |}
\hline
\bf Config & $F_\text{feat}$ & $F_\text{embed}$ & $F_\text{cls}$ & \bf \makecell{pre\\train}& \bf param loaded & \makecell{\bf freeze\\$F_\text{feat}$\\\bf 10epoch} & \bf \makecell{freeze\\$F_\text{embed}$\\\bf 10epoch} & \bf \makecell{input\\len\\(s)} & \bf loss & \bf DER\\
\hline
1  & Conv    & BLSTM & \multirow{4}{*}{linear} & \multirow{4}{*}{None}  & \multicolumn{3}{c |}{\multirow{4}{*}{-}} & \multirow{4}{*}{20} & \multirow{4}{*}{focal} & 0.486\\
2  & Conv    & MHA   &  &  & \multicolumn{3}{c |}{} & & & 0.490\\
3  & LogMel  & BLSTM &  &  & \multicolumn{3}{c |}{} & & & 0.535\\
4  & LogMel  & MHA   &  &  & \multicolumn{3}{c |}{} & & & 0.665\\
\hline
\textbf{5}  & Conv    & BLSTM & \multirow{4}{*}{\makecell{MLP\\2layer}} & \multirow{4}{*}{MIL1}  & $F_\text{feat}, F_\text{embed}$ & Yes & \multirow{4}{*}{Yes} & \multirow{4}{*}{20} & \multirow{4}{*}{focal} & \textbf{0.438}\\
6  & Conv    & MHA   & & & $F_\text{feat}, F_\text{embed}$ & Yes & & & & 0.509\\
7  & LogMel  & BLSTM & & & $F_\text{embed}$ & - & & & & 0.509\\
8  & LogMel  & MHA   & & & $F_\text{embed}$ & - & & & & 0.638\\
\hline
9  & Conv    & BLSTM & \multirow{4}{*}{\makecell{MLP\\2layer}} & \multirow{4}{*}{MIL1} & $F_\text{feat}, F_\text{embed}$ & No & \multirow{4}{*}{No} & \multirow{4}{*}{20} & \multirow{4}{*}{focal} & 0.461\\
10  & Conv    & MHA   & & & $F_\text{feat}, F_\text{embed}$ & No & & & & 0.446\\
11 & LogMel  & BLSTM & & & $F_\text{embed}$ & - & & & & 0.533\\
12 & LogMel  & MHA   & & & $F_\text{embed}$ & - & & & & 0.628\\
\hline
13  & Conv    & BLSTM & \multirow{4}{*}{linear} & \multirow{4}{*}{MIL2}  & $F_\text{feat}, F_\text{embed}, F_\text{cls}$ & Yes & \multirow{4}{*}{Yes} & \multirow{4}{*}{20} & \multirow{4}{*}{focal} & 0.465\\
14  & Conv    & MHA   & & & $F_\text{feat}, F_\text{embed}, F_\text{cls}$ & Yes & & & & 0.449\\
15  & LogMel  & BLSTM & & & $F_\text{embed}, F_\text{cls}$ & - & & & & 0.521\\
16  & LogMel  & MHA   & & & $F_\text{embed}, F_\text{cls}$ & - & & & & 0.656\\
\hline
17  & Conv    & BLSTM & \multirow{4}{*}{linear} & \multirow{4}{*}{MIL2} & $F_\text{feat}, F_\text{embed}, F_\text{cls}$ & No & \multirow{4}{*}{No} & \multirow{4}{*}{20} & \multirow{4}{*}{focal} & 0.514\\
18  & Conv    & MHA   & & & $F_\text{feat}, F_\text{embed}, F_\text{cls}$ & No & & & & 0.444\\
19 & LogMel  & BLSTM & & & $F_\text{embed}, F_\text{cls}$ & - & & & & 0.523\\
20 & LogMel  & MHA   & & & $F_\text{embed}, F_\text{cls}$ & - & & & & 0.631\\
\hline
LENA & \multicolumn{3}{c |}{-} & \multicolumn{4}{c |}{-} & - & - & {0.554}\\
\hline
\makecell{Lavechin\\et al. \cite{lavechin2020opensource}} & \makecell{Sinc-\\Net\cite{ravanelli2018speaker}} & BLSTM & \makecell{MLP\\3layer} & \multicolumn{4}{c |}{-} & 2 & MSE & {0.586}\\
\hline
\makecell{Ablation1} & Conv & BLSTM & linear & None & \multicolumn{3}{c |}{-} & 20 & BCE & 0.509\\
\makecell{Ablation2} & Conv & BLSTM & linear & None & \multicolumn{3}{c |}{-} & 2 & focal & 0.470\\
\hline
\makecell{Ablation3} & Conv & BLSTM & \makecell{MLP\\2layer} & MIL1  & $F_\text{feat}, F_\text{embed}$ & Yes & Yes & 2 & focal & 0.451\\
\hline

\end{tabular}
\caption{Performance of different system configurations}
\label{tab: results}
\end{table*}

\section{Pre-training with Multiple Instance Learning}
\label{sec: pretrain}

Because the time resolution of our diarization system is 256ms per frame, we require high-resolution labels to  train the system. However, infant recordings mostly consist of non-speech vocalizations, so the average speaker turns are extremely short \cite{7846253} compared with normal diarization datasets~\cite{EURECOM+3152}. In addition, overlapping vocalizations are frequent in infant recording. Therefore, it is relatively costly to acquire accurate labels; our core training and test datasets are precisely labeled, but relatively small.

Due to the limited size of our own training set, we use the Brawnwald~\cite{doi:10.1080/00437956.1971.11435613} and Providence~\cite{doi:10.1121/1.4795772} Corpora from the CHILDES project as a transfer learning dataset. During manual inspection, we noticed that in both datasets, although the labelled speaker turns are usually correct in terms of speaker class, turn boundaries are relatively imprecise: most speaker turns contain silence at both start and end.

Because we do not have the true binary speaker class label $\gY$ for each frame in our transfer learning dataset, we cannot directly train the same system described in \ref{sec: system}. Therefore, we re-formulate the problem as a multiple-instance learning problem, described below:

Given $(\rvx, s) \in \cD_{\text{pretrain}}$ as our transfer learning datum, where each pair of $\rvx$ and $0 < s < C$ are respectively the audio samples and the speaker class label for a segment, we wish to find a mapping $G_{\theta}$ that shares the same parameter $\theta$ with $F_\theta$. We wish to learn the parameters $\theta$ to maximize the accuracy of Eq.~\ref{eq:x2y}, but the true value of $\gY$ is unknown (time alignment of the utterance within the segment is unknown), therefore we design a classifier $G_\theta$ to maximize the accuracy of
\begin{equation}
\rvx \xrightarrow[]{G_{\theta}} \gY_{MIL} \in \left\{\mathbf{1}[s], \mathbf{0}\right\}
\label{eq:mil}
\end{equation}
where $\gY_{MIL}$, the multiple-instance learning target, is either $\mathbf{1}[s]$ (a one-hot vector for $s$) or $\mathbf{0}$ (the zero vector).
Eq.~\ref{eq:mil} can be  computed using a $G_\theta$ that computes the maximum over a segment, and compares the maximum to $\mathbf{1}[s]$, thus
\be \label{eq: MIL1}
G_\theta(\rvx)= (\text{SoftMax} \circ \text{MaxPool} \circ F_{\text{cls}} \circ F_{\text{embed}} \circ F_{\text{feat}})(\rvx)
\ee
or
\be \label{eq: MIL2}
G_\theta(\rvx) = (\text{Softmax} \circ F_{\text{cls}} \circ \text{MaxPool} \circ F_{\text{embed}} \circ F_{\text{feat}})(\rvx)
\ee
where $\text{MaxPool}$ denotes global max-pooling over time.

\section{Experimental Methods and Results}
\label{sec: experiment}

\label{ssec: lena}

Primary training, validation and test
data were drawn from two studies of socioemotional development. Families with typically developing infants between 3 and 24 months of age were recruited from the community. Infants wore the LENA recorder for a total of 16 hours in the home. 
Reference labels were manually coded for 107 10-minute segments that, according to LENA segmentation, had the highest frequency of voice activity (23 at 3 months, 20 at 6 months, 22 at 9 months, 22 at 12 months, and 20 at 13-24 months). We split those into 87 for train, 10 for validation, and 10 for test. For each 10-minute segment, we then manually labelled four tiers, with cross-labeler validation at a precision of 0.2 seconds: CHN (key child), CXN (other child), FAN (female adult), and MAN (male adult).
Neural nets were trained using audio waveforms normalized to $[-1, 1]$, and divided into 20-second segments with 0.256 second frames. We consider CHN and CXN as the same speaker class, and FAN and MAN as separate speaker classes.

\label{ssec: metrics}
We use Diarization Error Rate(DER)\cite{DER} as our primary metric. It can be computed as
\be
\frac{\sum_{s=1}^{S}\text{dur}(s) \cdot (\text{max}(N_{ref}(s), N_{hyp}(s)) - N_{correct}(s))}{\sum_{s=1}^{S}\text{dur}(s) \cdot N_{ref}}
\ee
Note that DER in the case of infant speech would be overall much higher than normal case, since the denominator of DER is smaller when less voice activity is present.

\label{ssec: pretrain}
Networks were pre-trained using the transfer learning datasets, 
Braunwald and Providence,
using two different MIL frameworks (MIL1=Eq.~\ref{eq: MIL1} and MIL2=Eq.~\ref{eq: MIL2}).
Variable-length segments were used, based on the start and end times in the labels. We only keep segments with durations between 1.28s
and 10.24s.
The best configurations using MIL1 and MIL2 achieved respectively 17.1\% and 16.7\% classification accuracy on validation set.

For both MIL1 and MIL2, we pre-train with all combinations of $F_\text{feat}$ and $F_\text{embed}$, and use a linear classifier with 3 classes(Child, Female, Male). We train with Adam optimizer for 5 epochs, with learning rate of 0.0005 and decay of 0.5 per epoch.

\label{ssec: train}
Configurations and results are listed in Table~\ref{tab: results}, using the notation  introduced in Sec.~\ref{sec: system}.  Configurations 1-4 and 9-12 usee learning rate 0.001 with decay of 0.98 per epoch. Configurations 5-8 used learning rate  0.0005 with decay 0.94 per epoch. Configurations 13-20 used learning rate 0.0005 with decay 0.98. 

\label{ssec: ablation}
We ran two baselines developed by others on our test set: LENA~\cite{Canault2015LENA,Cristia2018TalkerDI} and the system of Lavechin et al.~\cite{lavechin2020opensource}.
We ran three additional ablation studies on our baseline, each studying the effect of loss function, input segment size, and whether to use an additional speaker class for non-key child, each using the same learning rate as the configuration it ablates.

\label{ssec: results}
Results\footnote{code can be found at {\url{https://github.com/JunzheJosephZhu/Child_Speech_Diarization}}} are shown in \ref{tab: results}. Ablation studies 1, 2 and 3, compared to configurations 1, 1, and 5, respectively, show that focal loss improves performance and that shorter chunk size has uncertain effect on performance. Our best system is based on configuration 5, which achieves 43.8\% DER, compared to 55.4\% and 58.6\% DER achieved by baselines from LENA software and Lavechin et al.\cite{lavechin2020opensource}. We also note that convolutional feature extractors work better than logmel features in most cases, which contradicts prior practice in diarization systems. This could be due to the various forms of vocalizations in the infant speaker class, which include  extremely high-pitched vocalizations that may  be ill-adapted to logmel features.

Our best system (configuration 5) was also trained on a task with 4 speaker-class targets, in which the key child and other children are counted as separate classes.  DER and frame error rate (percentage of frames with prediction error in at least one class)  are reported in Table~\ref{tab:results2}. DER of LENA and our system were not affected much, but DER of \cite{lavechin2020opensource} increased.

\begin{table}[h]
\centering
\begin{tabular}{| c | c | c |}
\hline
\bf System & DER & Frame Error Rate\\
\hline
Ours, config 5 & 0.497 & 0.338\\
LENA & 0.581 & 0.353\\
Lavechin et al., 2020\cite{lavechin2020opensource} & 0.762 & 0.454\\
\hline

\end{tabular}
\caption{DER and Frame Error Rate of each system on 4-speaker case}
\label{tab:results2}
\end{table}

\section{Conclusions}
\label{sec: conclusions}
This paper offers two key contributions. First, we decomposed the child-speech diarization systems into three separately analyzed components: $F_\text{embed}$, $F_\text{feat}$, $F_\text{cls}$, and provided results for two configurations of each component. Second, we developed a pre-training procedure to enable transfer learning from datasets with coarse speaker segment labels. We found that convolutional features, combined with focal loss training and transfer learning, together achieves the most accurate system.



\vfill\pagebreak

\bibliographystyle{IEEEbib}
\bibliography{refs,mybib}

\end{document}